\documentclass[conference]{IEEEtran}
\IEEEoverridecommandlockouts
\usepackage{cite}
\usepackage{amsmath,amssymb,amsfonts}
\usepackage{algorithm}
\usepackage{algpseudocode}
\usepackage{graphicx}
\usepackage{textcomp}
\usepackage{xcolor}
\usepackage{booktabs}
\usepackage{tikz}
\usetikzlibrary{arrows.meta,positioning,calc,shapes.geometric}

\def\BibTeX{{\rm B\kern-.05em{\sc i\kern-.025em b}\kern-.08em
    T\kern-.1667em\lower.7ex\hbox{E}\kern-.125emX}}

\definecolor{boxfill}{rgb}{0.92,0.94,1.0}
\definecolor{boxedge}{rgb}{0.20,0.30,0.60}
\definecolor{accent}{rgb}{0.70,0.10,0.15}
\tikzset{
  blk/.style  = {rectangle, draw=boxedge, fill=boxfill, thick, align=center,
                 rounded corners=1.5pt, font=\scriptsize, inner sep=3pt,
                 minimum height=0.7cm},
  io/.style   = {rectangle, draw=gray, fill=white, thick, align=center,
                 font=\scriptsize, inner sep=3pt, minimum height=0.7cm},
  fl/.style   = {-{Latex[length=1.8mm]}, thick, draw=boxedge},
  tfl/.style  = {-{Latex[length=1.5mm]}, semithick, draw=gray}
}

\begin{document}

\title{AeroProp-X: An Open-Source, Self-Verifying\\
Method-of-Characteristics Toolkit for Edge-Capable\\
Rocket Nozzle Design and Aerothermodynamic Analysis}

\author{\IEEEauthorblockN{Aryuemaan Kumar Chowdhury}
\IEEEauthorblockA{\textit{Department of Lightweighting Engineering} \\
\textit{Indian Institute of Technology, Hyderabad}\\
Hyderabad, India \\
lw25mtech14004@iith.ac.in}
\IEEEauthorblockA{\textit{OSCOWL ai, Research and Development Department} \\
}}

\maketitle

\begin{abstract}
The design of supersonic rocket nozzles by the Method of Characteristics is a
mature discipline, yet the software that implements it is frequently either
proprietary, heavyweight, or opaque, which limits reproducibility and
classroom and field use. This paper presents AeroProp-X, an open-source,
dependency-light toolkit that synthesizes shock-free minimum-length nozzle
contours and evaluates their ideal aerothermodynamic performance entirely on
commodity hardware without graphics acceleration or network access. The central
contribution is methodological rather than physical: the planar
characteristics solver is made self-verifying by exploiting an exact
gas-dynamic invariant, namely that the computed exit wall height ratio must
equal the isentropic area ratio at the design Mach number, so that every
generated contour reports its own discretization error automatically. We
describe the governing theory, the layered software architecture, and the
synthesis algorithm, and we validate the implementation against analytic
compressible-flow tables and against the expansion ratios of historical flight
engines. The solver reproduces tabulated gas-dynamic quantities to table
precision and converges to the exact area ratio with relative error below three
parts in ten thousand, while remaining small enough to run interactively on an
embedded-class processor. We argue that built-in verifiability, dependency
minimalism, and explicit scoping of the ideal-flow assumptions together make
the tool well suited to education, preliminary design, and reproducible
research.
\end{abstract}


\section{Introduction}
The converging-diverging (de Laval) nozzle converts the thermal energy of a
high-pressure combustion gas into directed kinetic energy, and the shape of its
diverging wall determines how much of the available energy is recovered as
axial thrust. The reference method for synthesizing that wall is the Method of
Characteristics (MOC), which for steady, inviscid, irrotational supersonic flow
reduces the governing partial differential equation to ordinary relations along
characteristic curves and yields a contour that is simultaneously shock-free and
of minimum length \cite{anderson}.

Although the underlying theory is textbook material, accessible and trustworthy
implementations remain scarce. High-fidelity computational fluid dynamics is
accurate but expensive, slow, and ill-suited to the rapid trade studies of
early design or to teaching. Established propulsion analysis tools are powerful
but are often proprietary, tied to specific platforms, or distributed as
black boxes whose internal assumptions and numerical errors are difficult to
audit \cite{rpa,cea}. The result is a reproducibility gap: a student, a
hobbyist, or a researcher who wants a verifiable nozzle contour and an honest
estimate of its performance has few lightweight, inspectable options.

AeroProp-X addresses this gap. It is a compact, open-source toolkit that takes
chamber conditions and a target expansion and returns a characteristics-based
nozzle contour, its supersonic flow field, its ideal performance metrics, its
operating regime, and a solid-model export, using only a numerical and a
scientific-computing library. The contributions of this paper are:

\begin{itemize}
\item a \emph{self-verifying} MOC implementation, in which an exact
gas-dynamic invariant is used as a built-in correctness check so that each
contour reports its own discretization error;
\item a layered, dependency-light \emph{architecture} that runs on
embedded-class hardware without acceleration or network access, which we term
edge-capable;
\item a reproducible \emph{validation} of the solver against analytic tables
and historical engine geometries; and
\item an explicit, quantitative statement of the \emph{limits} of the
ideal-flow model, so that results are not over-trusted.
\end{itemize}

We emphasize at the outset that the physical models are classical; the novelty
lies in their integration, in the automatic verifiability of the solver, and in
the accessibility and honesty of the tool.

\section{Related Work}
Chemical-equilibrium performance codes such as NASA CEA compute idealized
specific impulse and exhaust composition for given propellants and pressures
\cite{cea}, but they do not synthesize a nozzle wall. Dedicated propulsion
analysis environments add geometry and one-dimensional performance
\cite{rpa}, yet they are typically closed and platform-specific. General CFD
solvers resolve viscous, reacting, separated flow at high cost and require
expertise and meshing effort disproportionate to preliminary design. The MOC
itself is developed thoroughly in classical references \cite{anderson,zucrow},
and thrust-optimized bell contours follow Rao's parabolic approximation
\cite{rao}; flow separation in over-expanded nozzles is commonly screened with
the empirical Summerfield criterion \cite{summerfield}. AeroProp-X differs from
these in combining contour synthesis, flow-field reconstruction, performance and
expansion analysis, and CAD export in a single inspectable, dependency-light
package whose solver verifies itself.

\section{Governing Theory}
The working fluid is treated as a calorically perfect gas with constant ratio
of specific heats $\gamma$ and specific gas constant $R$; the flow is steady,
adiabatic, inviscid, and shock-free, hence isentropic and irrotational.

\subsection{Quasi-one-dimensional relations}
Bringing the flow isentropically to rest defines the stagnation state, giving
the stagnation-to-static ratios used throughout,
\begin{equation}
\frac{T_0}{T}=1+\frac{\gamma-1}{2}M^2,\quad
\frac{p_0}{p}=\Bigl(\tfrac{T_0}{T}\Bigr)^{\!\frac{\gamma}{\gamma-1}},
\label{eq:stag}
\end{equation}
where $M$ is the Mach number. Mass conservation between a general station and
the sonic throat yields the area-Mach relation,
\begin{equation}
\frac{A}{A^\ast}=\frac{1}{M}\!\left[\frac{2}{\gamma+1}
\Bigl(1+\frac{\gamma-1}{2}M^2\Bigr)\right]^{\!\frac{\gamma+1}{2(\gamma-1)}},
\label{eq:area}
\end{equation}
which is double-valued; the diverging section takes the supersonic branch.
Equation~\eqref{eq:area} furnishes the key consistency relation exploited later:
a nozzle of exit Mach number $M_e$ must have area ratio
$\epsilon=A_e/A_t=A/A^\ast|_{M_e}$.

\subsection{Prandtl-Meyer expansion}
A supersonic stream turning through a convex angle expands isentropically. The
turning required to reach Mach number $M$ from sonic conditions is the
Prandtl-Meyer function,
\begin{equation}
\begin{split}
\nu(M)=\sqrt{\tfrac{\gamma+1}{\gamma-1}}\,
\arctan\!\sqrt{\tfrac{\gamma-1}{\gamma+1}\bigl(M^2-1\bigr)}\\
-\arctan\!\sqrt{M^2-1},
\end{split}
\label{eq:pm}
\end{equation}
with Mach angle $\mu=\arcsin(1/M)$. The inverse $M=\nu^{-1}(\nu)$ is obtained by
a guarded bracketed root find, since~\eqref{eq:pm} is smooth and monotonic.

\subsection{Characteristics of supersonic flow}
For two-dimensional irrotational supersonic flow the velocity potential $\phi$
obeys
\begin{equation}
\Bigl(1-\tfrac{u^2}{a^2}\Bigr)\phi_{xx}
-\tfrac{2uv}{a^2}\phi_{xy}
+\Bigl(1-\tfrac{v^2}{a^2}\Bigr)\phi_{yy}=0,
\label{eq:potential}
\end{equation}
which is hyperbolic when $M>1$. Its characteristics make angles $\pm\mu$ with
the streamline, with physical slopes $\tan(\theta\mp\mu)$, where $\theta$ is the
flow angle. Along them the flow satisfies the compatibility relations, which for
planar flow integrate to the Riemann invariants
\begin{equation}
\theta+\nu=J^{+}\ (\text{on }\Gamma^{-}),\qquad
\theta-\nu=J^{-}\ (\text{on }\Gamma^{+}).
\label{eq:inv}
\end{equation}
Knowing the two invariants at a point fixes $\theta$ and $\nu$, hence $M$,
$\mu$, and every thermodynamic ratio. For an axisymmetric nozzle a geometric
source term proportional to $\sin\theta/y$ augments~\eqref{eq:inv}; the present
solver is planar and omits it, a choice whose quantitative consequence is given
in Section~\ref{sec:disc}.

\section{System Architecture}
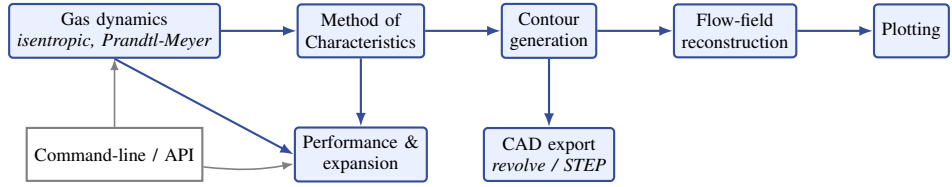
\begin{figure*}[t]
\centering
\begin{tikzpicture}[node distance=0.55cm and 1.0cm]
  \node[blk] (gas) {Gas dynamics\\\textit{isentropic, Prandtl-Meyer}};
  \node[blk, right=of gas] (moc) {Method of\\Characteristics};
  \node[blk, right=of moc] (noz) {Contour\\generation};
  \node[blk, right=of noz] (fld) {Flow-field\\reconstruction};
  \node[blk, right=of fld] (plt) {Plotting};
  \node[blk, below=0.9cm of noz] (cad) {CAD export\\\textit{revolve / STEP}};
  \node[blk, below=0.9cm of moc] (prf) {Performance \&\\expansion};
  \node[io,  below=0.9cm of gas] (cli) {Command-line / API};
  \draw[fl] (gas)--(moc); \draw[fl] (moc)--(noz);
  \draw[fl] (noz)--(fld); \draw[fl] (fld)--(plt);
  \draw[fl] (noz)--(cad); \draw[fl] (gas.south)--(prf.west);
  \draw[fl] (moc.south)--(prf.north);
  \draw[tfl] (cli)--(gas); \draw[tfl] (cli) to[bend right=8] (prf);
\end{tikzpicture}
\caption{Module architecture and principal data flow. Arrows point in the
direction of dependence and data movement. The gas-dynamics layer is shared by
both the characteristics solver and the performance model.}
\label{fig:arch}
\end{figure*}

AeroProp-X is organized as single-responsibility modules with a strict acyclic
dependency structure, shown in Fig.~\ref{fig:arch}. Low-level gas dynamics
underpins the characteristics solver; contour generation consumes the solver;
and the analysis, flow-field, visualization, and export layers sit on top. The
command interface orchestrates the lower layers but is depended upon by none of
them, which keeps the numerical core usable as a library and testable in
isolation.

\section{Self-Verifying Minimum-Length Nozzle Synthesis}
\label{sec:moc}
\subsection{Centred expansion at a sharp corner}
The minimum-length nozzle replaces the gradual throat arc of a conventional bell
with a single sharp corner that emits a centred Prandtl-Meyer fan. Requiring the
flow to expand to $M_e$ and then return to axial without compression fixes the
maximum wall angle at exactly half the exit Prandtl-Meyer angle,
\begin{equation}
\theta_{\max}=\tfrac12\,\nu(M_e).
\label{eq:thetamax}
\end{equation}
The corner is discretized into $n$ characteristics. Because every fan
characteristic originates at the sonic corner, each carries $J^{-}=0$, so on the
$i$-th ray $\nu_i=\theta_i$ and $J^{+}_i=2\theta_i$.

\subsection{Mesh construction}
\begin{figure}[t]
\centering
\begin{tikzpicture}[scale=0.92]
  \draw[gray,dashed] (-0.1,0)--(5.0,0) node[right,font=\scriptsize]{axis};
  \fill[accent] (0,1.45) circle (1.6pt) node[left,font=\scriptsize]{corner};
  \foreach \a in {1.45,1.05,0.65,0.28,0.03}{\draw[accent,very thin] (0,1.45)--(4.8,\a);}
  \draw[boxedge,very thin] (1.25,0)--(2.7,1.05);
  \draw[boxedge,very thin] (2.15,0)--(3.5,0.9);
  \draw[boxedge,very thin] (3.0,0)--(4.2,0.78);
  \draw[very thick,boxedge] (0,1.45) .. controls (1.5,1.95) and (3.2,1.85)
        .. (4.8,1.72) node[above right,font=\scriptsize]{wall};
  \draw[teal,thick,dashed] (4.8,0)--(4.8,1.72);
  \node[teal,font=\scriptsize] at (4.6,0.3) {exit};
  \node[accent,font=\scriptsize] at (1.2,0.5) {$\Gamma^{-}$ fan};
  \node[boxedge,font=\scriptsize] at (2.5,1.25) {$\Gamma^{+}$};
\end{tikzpicture}
\caption{Characteristic mesh of a planar minimum-length nozzle. The sharp corner
emits a centred $\Gamma^{-}$ fan; reflected $\Gamma^{+}$ characteristics are
cancelled along the wall, giving a shock-free contour and a uniform, axial
exit.}
\label{fig:mesh}
\end{figure}
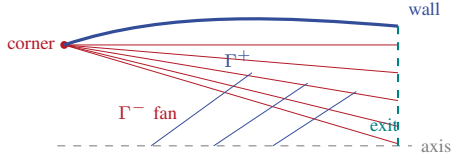

The solution is built as a triangular array of nodes $P(i,j)$, $i=1,\dots,n$,
$j=1,\dots,i$. Using the corner invariants and the unit processes of
Fig.~\ref{fig:mesh}, the interior and axis states take the closed form
\begin{equation}
\begin{aligned}
\text{interior }(j<i):&\quad \theta_{i,j}=\theta_i-\theta_j,\ \
\nu_{i,j}=\theta_i+\theta_j,\\
\text{axis }(j=i):&\quad \theta_{i,i}=0,\ \ \nu_{i,i}=2\theta_i,
\end{aligned}
\label{eq:mesh}
\end{equation}
which satisfy~\eqref{eq:inv} identically. Each interior node is the intersection
of an incoming $\Gamma^{+}$ (carrying $J^{-}$ from node~1) and $\Gamma^{-}$
(carrying $J^{+}$ from node~2),
\begin{equation}
\theta_3=\tfrac12\bigl(J^{+}_2+J^{-}_1\bigr),\qquad
\nu_3=\tfrac12\bigl(J^{+}_2-J^{-}_1\bigr),
\label{eq:interior}
\end{equation}
and its position follows from intersecting the two characteristic segments using
slopes averaged between the known and new nodes, a second-order predictor. Wall
nodes are placed by cancelling each reflected $\Gamma^{+}$ at the wall
streamline, which makes the contour shock-free and of minimum length.
Algorithm~\ref{alg:moc} summarizes the synthesis.

\begin{algorithm}[t]
\caption{Planar minimum-length nozzle synthesis}
\label{alg:moc}
\begin{algorithmic}[1]
\Require exit Mach $M_e$, ratio $\gamma$, count $n$, throat half-height $y_t$
\State $\theta_{\max}\gets\tfrac12\nu(M_e)$ \Comment{Eq.~\eqref{eq:thetamax}}
\State distribute $0<\theta_1<\dots<\theta_n=\theta_{\max}$
\For{$i=1$ to $n$}
  \For{$j=1$ to $i$}
    \State set $\theta_{i,j},\nu_{i,j}$ by Eq.~\eqref{eq:mesh}
    \State $M_{i,j}\gets\nu^{-1}(\nu_{i,j})$;\ \ $\mu_{i,j}\gets\arcsin(1/M_{i,j})$
    \State locate $(x_{i,j},y_{i,j})$ by averaged-slope intersection
  \EndFor
\EndFor
\State build wall by cancelling reflected $\Gamma^{+}$ at the wall
\State scale coordinates by $y_t$
\State \textbf{return} node states, wall polyline, and error $\varepsilon_A$
\end{algorithmic}
\end{algorithm}

\subsection{The self-verification invariant}
The methodological core of the tool is that the solver checks itself. For planar
flow the exit wall height ratio must equal the isentropic area
ratio~\eqref{eq:area} at the design Mach number. The solver therefore reports,
for every contour, the relative discrepancy
\begin{equation}
\varepsilon_A=\left|\frac{(y_e/y_t)_{\mathrm{MOC}}}{A/A^\ast|_{M_e}}-1\right|,
\label{eq:eps}
\end{equation}
which is an \emph{exact} consequence of the physics, not a heuristic. Two further
invariants hold at every resolution and are also checked: the peak centreline
Mach number equals $M_e$, and the exit wall flow angle returns to zero. Together
these turn correctness from something asserted into something measured: a
malformed mesh, an indexing error, or an under-resolved fan all manifest as a
non-vanishing $\varepsilon_A$. This automatic verifiability is, to our
knowledge, not surfaced by comparable lightweight tools.

\section{Performance and Expansion Modeling}
Once the geometry and operating pressures are known, the ideal one-dimensional
metrics follow in closed form. Introducing the Vandenkerckhove function
\begin{equation}
\Gamma(\gamma)=\sqrt{\gamma}\,
\Bigl(\tfrac{2}{\gamma+1}\Bigr)^{\!\frac{\gamma+1}{2(\gamma-1)}},
\label{eq:vk}
\end{equation}
the characteristic velocity, thrust coefficient, and exhaust velocity compact to
\begin{align}
c^\ast&=\frac{\sqrt{R\,T_0}}{\Gamma},\label{eq:cstar}\\
C_F&=\Gamma\sqrt{\frac{2\gamma}{\gamma-1}
\Bigl[1-\bigl(\tfrac{p_e}{p_0}\bigr)^{\frac{\gamma-1}{\gamma}}\Bigr]}
+\epsilon\,\frac{p_e-p_a}{p_0},\label{eq:cf}\\
v_e&=\sqrt{\frac{2\gamma}{\gamma-1}R\,T_0
\Bigl[1-\bigl(\tfrac{p_e}{p_0}\bigr)^{\frac{\gamma-1}{\gamma}}\Bigr]},
\label{eq:ve}
\end{align}
with chamber pressure $p_0$, exit pressure $p_e$, ambient pressure $p_a$, and
area ratio $\epsilon$. The specific impulse and thrust are
$I_{sp}=C_F c^\ast/g_0$ and $F=C_F p_0 A_t$, with
$g_0=9.80665~\mathrm{m\,s^{-2}}$. The first term of~\eqref{eq:cf} is the
momentum contribution and the second the pressure-thrust correction, positive
when under-expanded and negative when over-expanded.

The operating regime is set by comparing $p_e$ with $p_a$: the flow is
perfectly expanded when $p_e\approx p_a$, under-expanded when $p_e>p_a$ (vacuum
being the limit), and over-expanded when $p_e<p_a$. Strong over-expansion can
separate the boundary layer; this is screened with the Summerfield
criterion, which anticipates separation when
\begin{equation}
\frac{p_e}{p_a}<\kappa_{\mathrm{sep}},\qquad
\kappa_{\mathrm{sep}}\approx0.35\text{--}0.40.
\label{eq:sep}
\end{equation}
A conservative threshold of $0.40$ is used by default; the criterion predicts
whether separation is likely, not where it occurs nor the resulting side loads.

\section{Results and Validation}
\subsection{Gas-dynamic relations}
Table~\ref{tab:isentropic} compares computed isentropic and Prandtl-Meyer
quantities against standard compressible-flow tables for $\gamma=1.4$. Agreement
is to table precision across the Mach range.

\begin{table}[t]
\caption{Gas-dynamic quantities: computed versus tables ($\gamma=1.4$)}
\label{tab:isentropic}
\centering
\footnotesize
\begin{tabular}{ccccc}
\toprule
$M$ & $A/A^\ast$ code & $A/A^\ast$ tab. & $\nu$ code & $\nu$ tab.\\
\midrule
2.0 & 1.6875 & 1.6875 & $26.380^\circ$ & $26.380^\circ$\\
2.4 & 2.4031 & 2.4031 & $36.747^\circ$ & $36.747^\circ$\\
3.0 & 4.2346 & 4.2346 & $49.757^\circ$ & $49.757^\circ$\\
4.0 & 10.719 & 10.719 & $65.785^\circ$ & $65.785^\circ$\\
\bottomrule
\end{tabular}
\end{table}

\subsection{Convergence of the characteristics solver}
\begin{figure}[t]
\centering
\includegraphics[width=0.86\columnwidth]{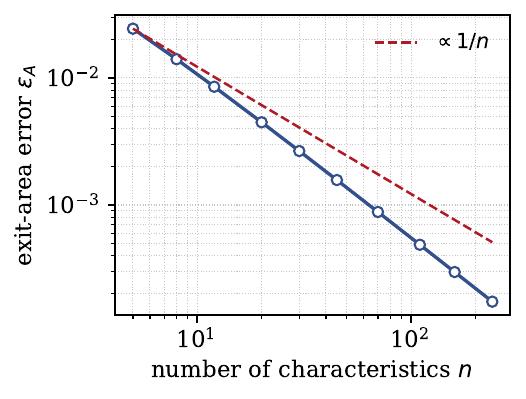}
\caption{Self-verification error $\varepsilon_A$ versus characteristic count at
$M_e=2.4$. The dashed line shows a first-order $1/n$ reference.}
\label{fig:conv}
\end{figure}

For the design point $M_e=2.4$ ($\gamma=1.4$, exact $A/A^\ast=2.4031$,
$\theta_{\max}=18.37^\circ$), the self-verification error~\eqref{eq:eps} falls
monotonically under mesh refinement, as listed in Table~\ref{tab:conv} and
plotted in Fig.~\ref{fig:conv}. The error decreases approximately as $1/n$ and
reaches below $3\times10^{-4}$ at $n=160$, confirming both convergence to the
exact area ratio and the correctness of the implementation.

\begin{table}[t]
\caption{MOC convergence at $M_e=2.4$}
\label{tab:conv}
\centering
\footnotesize
\begin{tabular}{ccc}
\toprule
$n$ & exit height ratio & error $\varepsilon_A$\\
\midrule
5   & 2.3444 & $2.44\times10^{-2}$\\
20  & 2.3923 & $4.48\times10^{-3}$\\
40  & 2.3987 & $1.83\times10^{-3}$\\
80  & 2.4013 & $7.40\times10^{-4}$\\
160 & 2.4024 & $2.97\times10^{-4}$\\
\bottomrule
\end{tabular}
\end{table}

\subsection{Reconstructed flow field}
\begin{figure}[t]
\centering
\includegraphics[width=0.96\columnwidth]{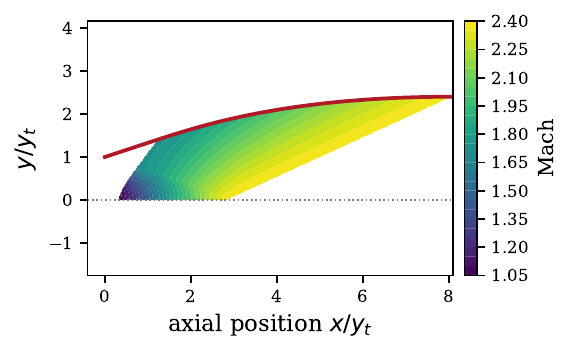}
\caption{Mach-number field of a planar minimum-length nozzle ($M_e=2.4$,
$n=70$) reconstructed from the characteristic mesh; the wall contour is shown in
red. Coordinates are normalized by the throat half-height.}
\label{fig:field}
\end{figure}

Fig.~\ref{fig:field} shows the Mach field reconstructed from the characteristic
mesh for an $M_e=2.4$ nozzle ($n=70$). The flow accelerates monotonically from
just above sonic at the throat corner to the design Mach number, which is
attained uniformly across the exit plane, and the wall returns the flow to
axial as required.

\subsection{Comparison with flight hardware}
Table~\ref{tab:engines} lists representative expansion ratios of flight engines;
the contour generators reproduce these geometric families. The comparison is
geometric by intent: published engines employ real-gas and finite-rate
chemistry, film cooling, and regenerative passages that the ideal model does not
capture, so contour shape and expansion ratio, not exact specific impulse, are
the meaningful points of comparison. The deliberately ideal specific impulse
computed by AeroProp-X therefore lies below flight values, and that gap is itself
consistent with the stated scope of the model.

\begin{table}[t]
\caption{Representative engine expansion ratios (illustrative)}
\label{tab:engines}
\centering
\footnotesize
\begin{tabular}{lcc}
\toprule
Engine & $A_e/A_t$ & AeroProp-X family\\
\midrule
RS-25 (sea-level) & $\approx 69$ & Rao 80\% bell\\
Vulcain 2 & $\approx 60$ & Rao 80\% bell\\
Vikas (vacuum) & $\approx 31$ & Rao 80\% bell\\
\bottomrule
\end{tabular}
\end{table}

\subsection{Footprint}
The toolkit depends only on a numerical array library and a scientific-computing
library; it requires no graphics acceleration, no compiled extensions, and no
network access at run time. A full minimum-length synthesis with several tens of
characteristics, together with performance evaluation, completes in well under a
second on a commodity processor and within the memory budget of an
embedded-class device, supporting the edge-capable claim.

\section{Discussion}\label{sec:disc}
The principal limitation is that the characteristics solver is planar. Revolving
its meridional contour as an axisymmetric body squares the height ratio and thus
overshoots the design area ratio by a factor $\sqrt{A/A^\ast}$; consequently the
generator defaults to the planar interpretation, where~\eqref{eq:eps} holds
exactly, and the Rao parabolic bell is provided for accurate axisymmetric area
ratios. A rigorous axisymmetric MOC, incorporating the $\sin\theta/y$ source
term, is the foremost item of future work. More broadly, the inviscid,
calorically perfect, shock-free model excludes boundary layers, chemistry, and
heat transfer, so the performance figures are ideal upper estimates; the Rao
default angles and the Summerfield threshold are likewise engineering
approximations. We regard the explicit, quantitative disclosure of these limits
as a feature: it is what allows the tool to be trusted within its domain and not
beyond it. The self-verification invariant of Section~\ref{sec:moc} guards only
the gas-dynamic correctness of the planar solver; it does not, and is not
claimed to, certify the physical fidelity of the underlying assumptions.

\section{Conclusion}
We presented AeroProp-X, an open-source, dependency-light, edge-capable toolkit
for rocket nozzle design and aerothermodynamic analysis built around a planar
Method-of-Characteristics solver. Its distinguishing feature is
self-verification: an exact gas-dynamic invariant lets every synthesized contour
report its own discretization error, turning correctness into a measured
quantity. The implementation reproduces analytic gas-dynamic tables to table
precision, converges to the exact area ratio with relative error below three
parts in ten thousand, and reproduces the expansion-ratio families of flight
engines, all within a footprint small enough for interactive use on commodity or
embedded hardware. By combining verifiability, accessibility, and an honest
statement of its ideal-flow scope, the tool aims to serve education,
preliminary design, and reproducible research. Future work targets a rigorous
axisymmetric solver and an optional boundary-layer correction.


\end{document}